\documentclass[prb]{revtex4}

\usepackage{graphicx}
\usepackage{dcolumn}
\usepackage{bm}

\begin{document}


\title{Joint analysis of spectral reactor neutrino experiments}

\author{V.V. Sinev}
\affiliation{%
Institute for Nuclear Research RAS, Moscow\\
}%


\begin{abstract}
The analysis of experiments at nuclear reactors where inverse beta decay reaction positron spectrum was measured at different distances from reactor core is presented here. It was found that there appear two closed zones of neutrino oscillation parameters when joint analysis is applied on the ${\Delta}m^{2}-\sin^{2}2{\theta}$ plane. The parameters that found are partially crossed with similar regions originating from other non reactor experiments where they observed neutrino oscillations having unusual mass parameter $\sim$1 eV$^2$ and amplitude about 0.08. Confidence level for observed regions achieves the value of 99.9\%. 
\end{abstract}

\maketitle

\section*{Introduction}

At the end of twentieth century there were done several experiments on looking for neutrino oscillations at nuclear reactors all over the world. They were stimulated by the lack of events in solar Cl-Ar experiment made by Davis et al. [1]. The same lack was observed in latter experiment with Ga [2]. Oscillation parameters were unknown that times and reactor experiments were done at a number of distances in hope to find some oscillations similar to solar ones. That time they did not observe oscillations at reactors.

Evidence for neutrino oscillations was clearly appeared in experiment Kamiokande [3]. They registered atmospheric neutrinos and found anomaly in muon neutrino events registration. Later it was proved with more powerful experiment Super Kamiokande [4]. A little bit later the evidence of solar neutrinos oscillations was proved by experiments SNO [5] and KamLAND [6]. They appeared to be the same as predicted by theory of Micheev-Smirnov-Wolfenstein [7]. SNO proved the predicted boron neutrino flux accounting interactions of muon and taon neutrinos in the target that cannot be seen in electron neutrino interactions. KamLAND measured the distortion of reactor antineutrino spectrum caused by the same parameters as solar neutrinos demonstrate. 

Experiments with solar and atmospheric neutrinos give mass oscillation parameters that differ on two degrees. That proves three active neutrino existence. Analysis of Z$^0$ bozon decay leads to the number of active neutrinos equals to 3, what means that there are to exist three mass square differences. But in case of large difference in their value it is enough to have only two mass differences. Exactly this one can see in reality: ${\Delta}m^{2}_{atm}=2.4\times 10^{-3}$ eV$^2$, but ${\Delta}m^{2}_{sol}=7.9\times 10^{-5}$ eV$^2$.

A little bit aside there are experiments that found some specific neutrino oscillations with another mass parameter differing from both mentioned above. They are LSND [8] experiment and MiniBOONE [9]. They demonstrate a region of parameters that are not proved to the moment by other experiments. KARMEN [10] $-$ experiment very similar to MiniBOONE does not observe the same effect as MiniBOONE.

Last time there appeared two more evidences of unusual neutrino oscillations with similar parameters as LSND and MinBOONE. First of them is result of calibration Ga-Ge experiment that was done on both setups SAGE and GALLEX [11] with $^{51}$Cr and $^{37}$Ar. In both experiments they observed lack of events. Second is the last calculation of antineutrino reactor flux [12] that appeared 3\% higher than used before. Authors reanalyze data on measuring reactor flux by the number of experiments and found that mean value is 0.94 instead 0.98. They regard this as evidence of oscillation and calculated neutrino oscillation parameters regions [13]. 

It is necessary to mention that in reactor experiments they found neutrino oscillation parameters that satisfied the experimental data but they were out of experiments sensitivity. In 2000 a special experiment [14] to check these parameters was proposed because their values appeared to be very close each other. 

In this paper the joint analysis of 6 reactor experiments is presented. All these experiments used inverse beta decay reaction for neutrino registration and measured energy spectrum of positrons from the reaction.
\begin{equation}
\bar{\nu_{e}}+p \rightarrow n + e^{+}.
\end{equation}

In table 2 one can find some experiment parameters used in presented analysis. Last two rows of the table contain experiments that measured the spectrum at only one distance and compare it with expected one.

\begin{table}[h]
\caption{the chi squared function minimum and neutrino oscillation parameters found in analyzing experiments.}
\label{table:1}
\vspace{10pt}
\begin{tabular}{l|c|c|c|c|c}
\hline
Site &Distance to core, m & ${\Delta}m^{2}$, eV$^2$ & $\sin^{2}2{\theta}$ & ${\chi}^2_{min}$ & cite \\
\hline
Gosgen& 37.9 45.9 64.7 & 0.88 & 0.1$-$0.06 & 38.8 & [15] \\
Rovno & 18.2 25.2 & 0.9 & 0.09 & 27 & [16] \\
Bugey-3& 15 40 95 & 0.45$-$1.7 & 0.09$-$0.03  & 33 & [17] \\
SRP & 18.2 23.8 & 3.84 & 0.085 & 12.6  & [18] \\
ILL & 8.76 & 2.23 & 0.31 & $-$  & [19] \\
Rovno-2& 18.2 & 0.9 & 0.1 & $-$  & [20] \\
\hline
\end{tabular}\\[2pt]
\end{table} 

\section{The fit for experimental positron spectra}

When analyzing experimental data it is very important to know the detector response function to be able describe correctly experimental spectra. The detector response function is the spectrum observed in a detector when registering positron with known energy.

In some papers one can find detector response functions. At figure 1 response functions for Gosgen [14] and Rovno [15] detectors are shown. The features of these functions are two parts: Gaussian and tail going to zero. Gaussian has origin from the positron kinetic energy and tail comes from some rest energy leaving by annihilation gammas escaping from fiducial volume and positrons appeared near the detector edge. Gammas in organic scintillators usually are registered through a number of Compton scattering.

In different detectors and experiments gauss position depends on additional energy from annihilation gammas. It may be possible to use some mean response function fitting each experiment to analyze all experiments jointly. At fig. 2 one can see how using the same response function but shifting on the energy of annihilation gammas released energy we fit analyzing positron spectra. Positron spectrum can be written as:
\begin{equation}
S(E_{e})=\int{f_{\nu}(E_{\nu}){\sigma}_{0}(E_{\nu})(1+{\delta}(E_{\nu})R(E_{e},T_{e})dE_{\nu}dT_{e}},
\end{equation}
where ${\sigma}_{0}(E_{\nu})$ $-$ zero approximation cross section of reaction (1) and ${\delta}(E_{\nu})$ $-$ correction connected with recoil, weak magnetism and radiation. $R(E_{e},T_{e})$ $-$ detector response function containing $T_{e}$ $–$ positron kinetic energy and $E_{e}$ $–$ energy registered in the detector.

\begin{figure}
\includegraphics[width=100mm]{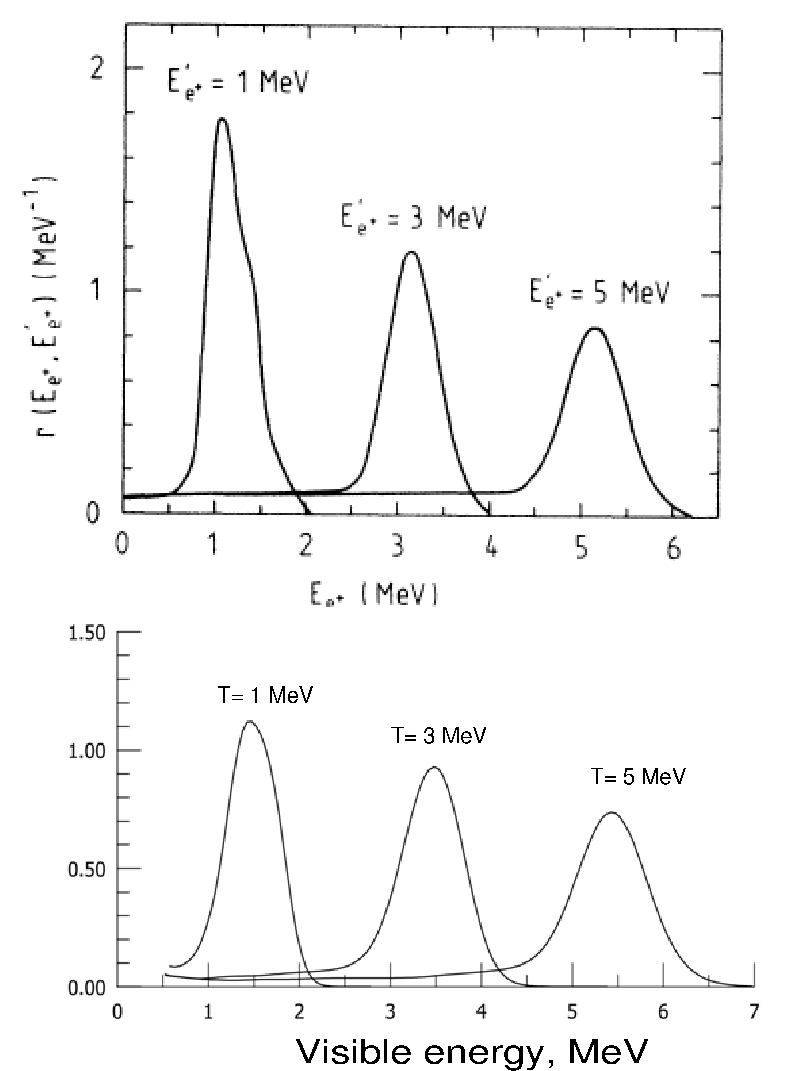}
\caption{\label{fig:fig1} Response function for detectors at Gosgen (upper panel) and Rovno (lower panel).}
\end{figure}

\begin{figure}
\includegraphics[width=100mm]{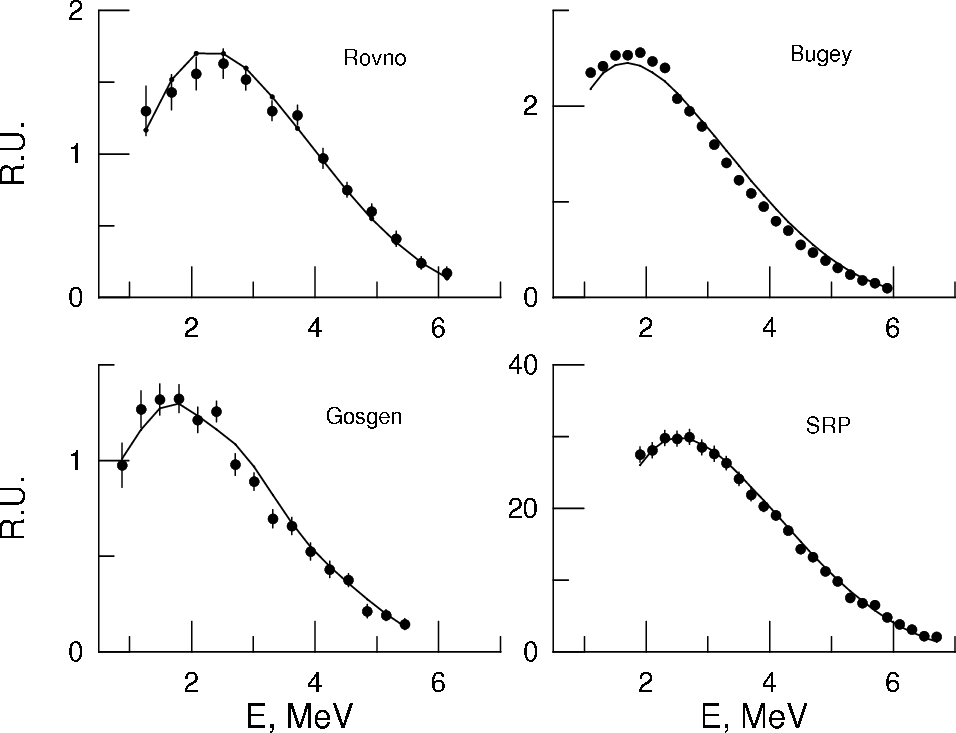}
\caption{\label{fig:fig1} Experimental positron spectra and fitted spectra calculated on base of universal response function.}
\end{figure}

\section{Analysis of spectra shape}

To analyze the shape of positron spectra we will regard spectra ratios. Ratios will be constructed only in ranges of separate experiment. If experimental ratio of spectra made at different distances was published we took it from corresponding paper if not we calculate them ourselves. Some experimental ratios with best fits are shown at figure 3.

\begin{figure}
\includegraphics[width=100mm]{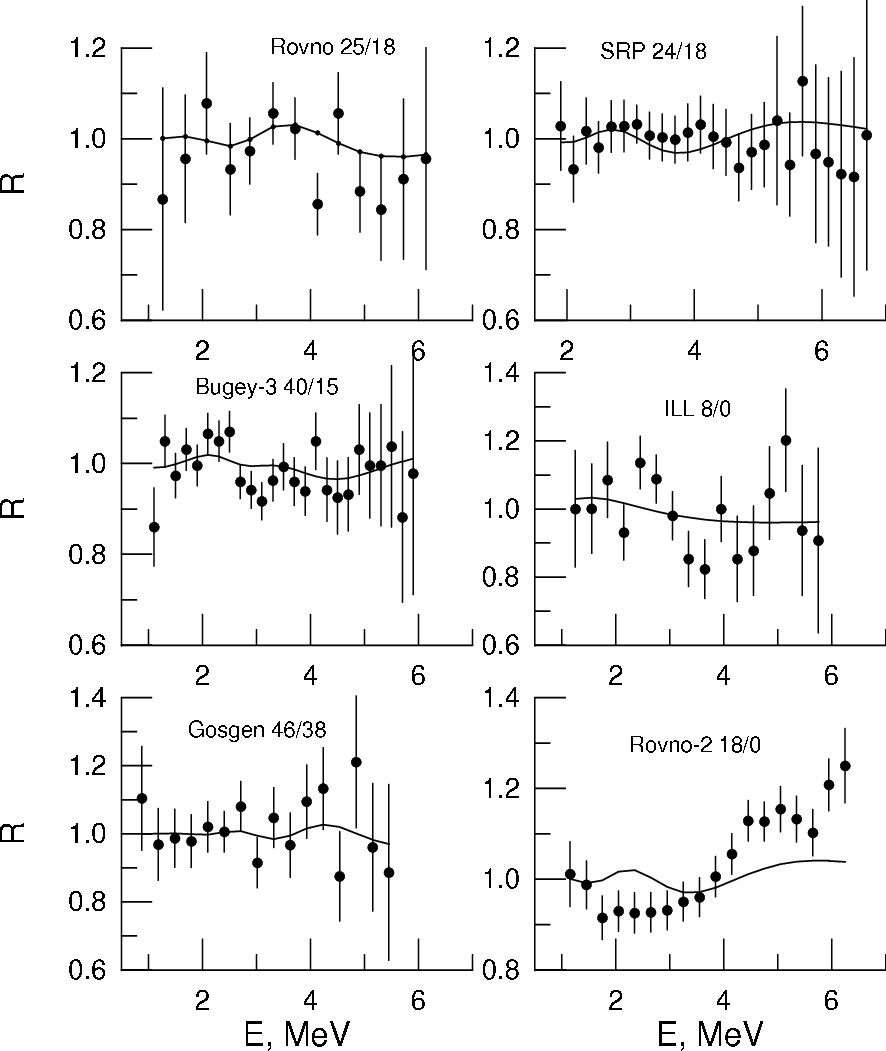}
\caption{\label{fig:fig1} Experimental positron spectra ratios and best fit.}
\end{figure}

For the analysis we used chi-square function organized like this

\begin{equation}
{\chi}^2=\sum_{k}\sum_{i}{\frac{(^{exp}R_{ik}-^{calc}R_{ik})^2}{{\sigma}^2_{ik}}},
\end{equation}
where $^{exp}R$ $-$ experimental ratio, $^{calc}R$ $-$ fitted ratio for choosen ${\Delta}m^2$ and $\sin^{2}2\theta$ and ${\sigma}^2_{ik}$ $-$ standard deviation. Index $i$ stands for spectrum bin and $k$ for the experiment.

Analysis was done in several stages.
Analysis A. There were analyzed only experiments where several distances were used for positron spectra measurement. Minimal value of ${\chi}^2$ was found ${\chi}^2_{min}=64.17$ at ${\Delta}m^2$=0.98 eV$^2$ and $\sin^{2}2\theta$=0.05. ${\chi}^2(0,0)=68.62$ in absence of oscillation. Total bins number is 121 (Rovno $-$ 13, Gosgen $-$ 3õ16, Bugey $-$ 25+10, SRP $-$ 25). If to look for the curve at 90\% CL (4.61 for 2 dof ${\chi}^2$ distribution) it comes open, but 89\% CL (4.41) encloses regions of possible parameters. See figure 4.  

\begin{figure}
\includegraphics[width=100mm]{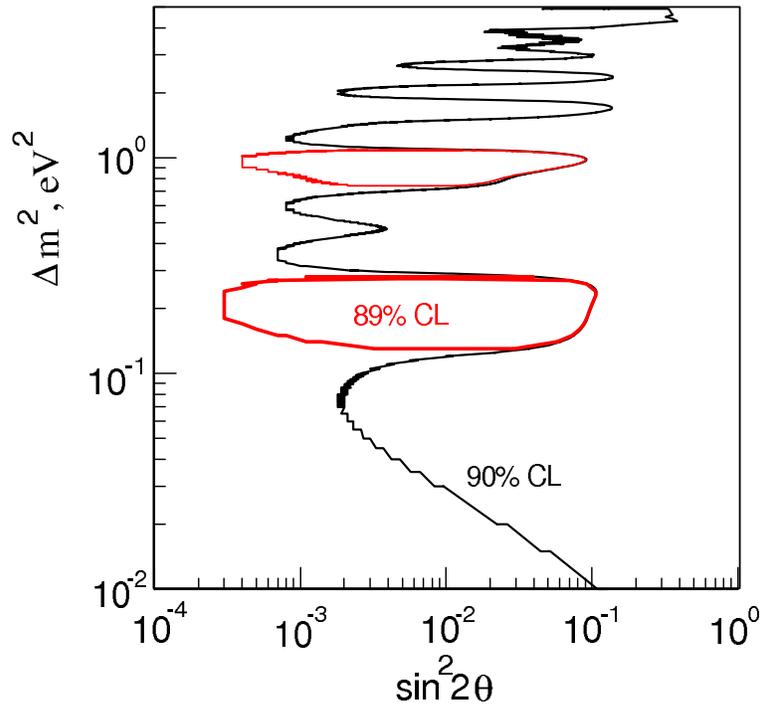}
\caption{\label{fig:fig1} 90\% limitation contour for joint analysis of experiments [15],[16],[17],[18].}
\end{figure}

It is understandable because the difference between ${\chi}^2(0,0)$ and ${\chi}^2_{min}$ is less than 4.61 but more than 4.41.

Analysis B. To Analysis A was added experiment at ILL [19]. ${\chi}^2_{min}=88.18$ at ${\Delta}m^2$=0.97 eV$^2$ and $\sin^{2}2\theta$=0.046. ${\chi}^2(0,0)=94.14$ in absence of oscillation. Total bins number is 137. There appeared 4 closed areas around 0.2, 1, 2.2 and 3.8 eV$^2$ at 90\% CL. But beginning from 95\% CL curves are still open (see figure 5).

\begin{figure}
\includegraphics[width=100mm]{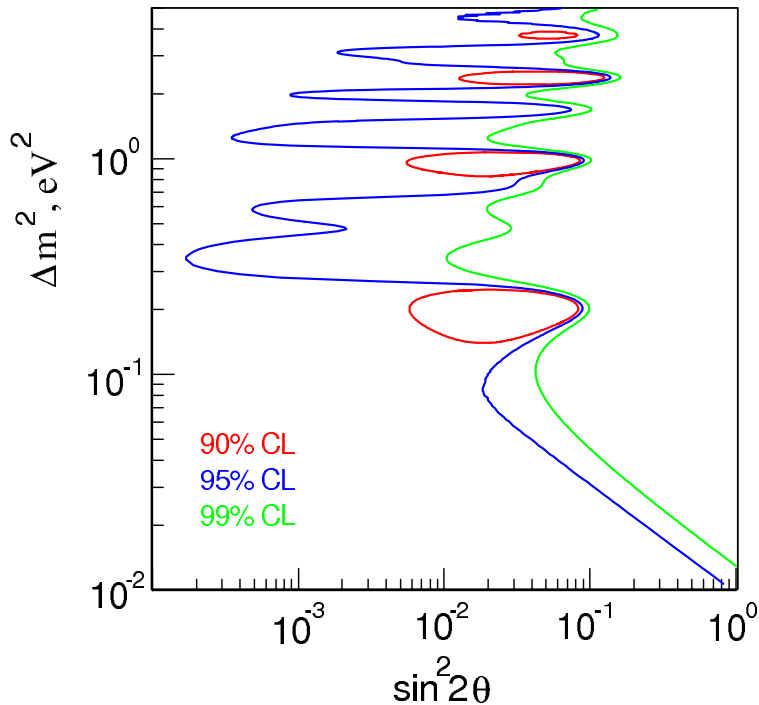}
\caption{\label{fig:fig1} 90\%, 95\% and 99\% limitation contours for joint analysis of experiments [15],[16],[17],[18] and [19].}
\end{figure}

Analysis C. In this Analysis we add high statistics experiment made at Rovno in 1988-1990 [20]. We call it Rovno-2. The same tipe detector was installed at a distance of 18 m to reactor core instead of previuos one. It is volume was a factor 2 larger and in addition surrounded with a layer of scintillator for capturing with higher efficiency annihilation gammas. Dayly statistics was grown in 3 times. Unfortunately there was no ability to use position at 25 m. So, the comparison could be done only with expected spectrum. There appeared two areas enclosed with a curves 95\% and 99\%, but only one for 90\% CL.  ${\chi}^2_{min}=144.57$ at ${\Delta}m^2$=0.208 eV$^2$ and $\sin^{2}2\theta$=0.082. ${\chi}^2(0,0)=166.35$ in absence of oscillation. Total bins number is 155. The difference is more than 4 standard deviations for gaussian distribution.
We used distribution $e^{-frac{{\chi}^2}{2}}$ to check how sharp is our experimental ${\chi}^2$ distribution. At figure 6 (upper panel) the ${\chi}^2$ values shown in vicinity of the minimum and at lower panel $e^{-frac{{\chi}^2}{2}}$ distribution.

\begin{figure}
\includegraphics[width=100mm]{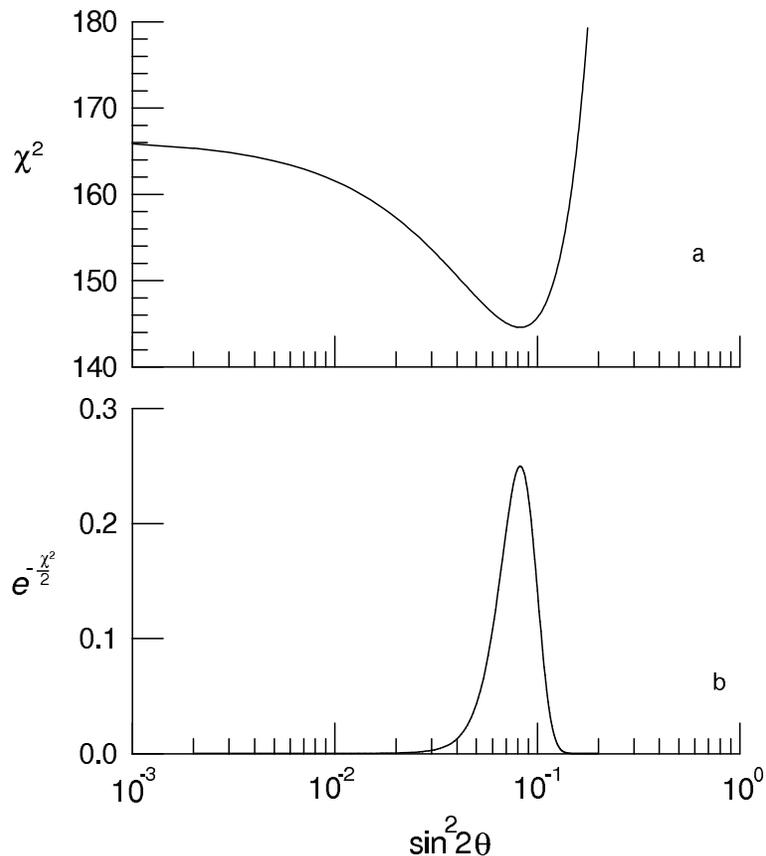}
\caption{\label{fig:fig6} (Upper panel) ${\chi}^2$ in vicinity of minimum. Cross section through ${\Delta}m^2$=0.925 eV$^2$. (Lower panel) $e^{-frac{{\chi}^2}{2}}$ distribution.}
\end{figure}

At figure 7 we compare found in presented analysis oscillation parameters areas with areas found in analysis of other experiments ([9],[11] and [17]). It is appeared that new parameters are in the region forbidden with Bugey-3. That is because we do not account the normalization of spectra, only the shape was analyzed. We can get the limit on the value of $\sin^{2}2\theta$ if add to the analysis ratios of measured to expected cross sections for all reactor neutrino experiments. They are shown at figure 8.   

\begin{figure}
\includegraphics[width=100mm]{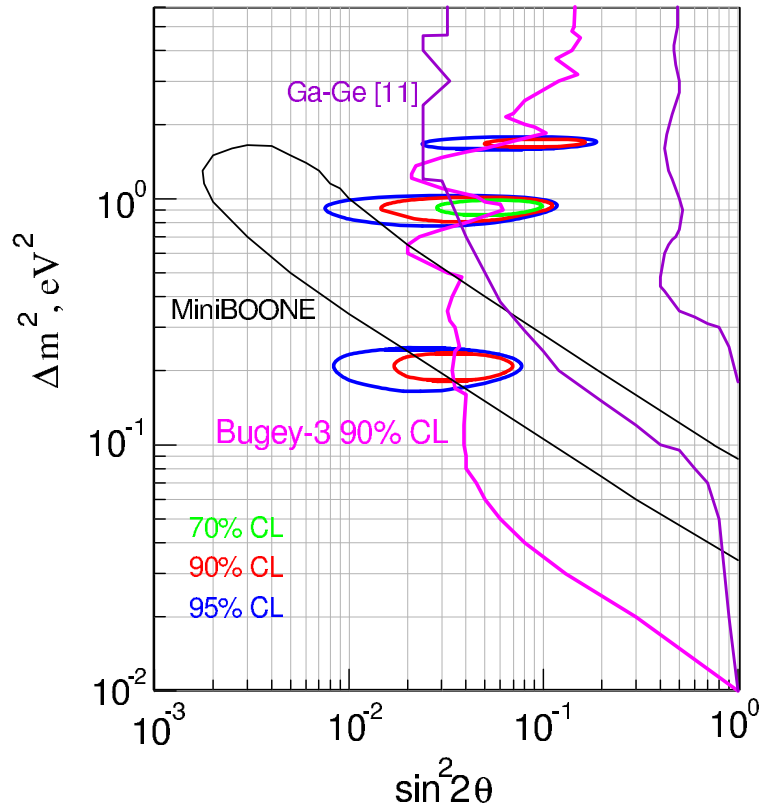}
\caption{\label{fig:fig7} 90\%, 95\% and 99\% limitation contours for joint analysis of experiments [15],[16],[17],[18] and [19] and 90\% contours for [9], [11] and [17]. Black points mark locations of minima.}
\end{figure}

\begin{figure}
\includegraphics[width=100mm]{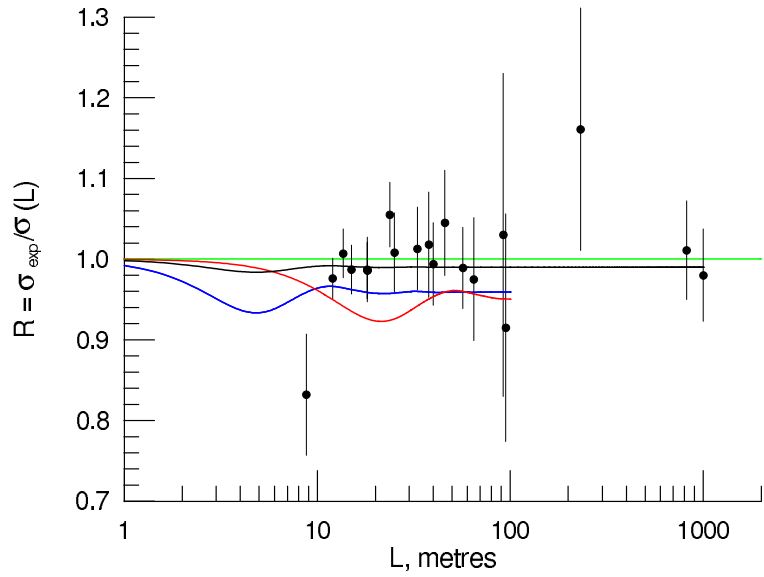}
\caption{\label{fig:fig1} Ratio of experimental cross sections to expected in absence of oscillation ones for most of reactor neutrino experiments.Blue and red curves are calculated for parameters obtained from our found areas. Black curve is calculated for parameters: ${\Delta}m^2$=0.925 eV$^2$ and $\sin^{2}2\theta$ = 0.02.}
\end{figure}
 
Black curve calculated for mass parameter and $\sin^{2}2\theta$ = 0.02 found in our analysis satisfies to all experiments at reactors. 
In [13] authors used their spectrum to ``correct'' ratios. In case of really existing reactor spectrum close to their esimation our both parameters might satisfy to flux ratios. In table II one can find ratios of measured to expected fluxes used at the figure 8.

\begin{table}[h]
\caption{the chi squared function minimum and neutrino oscillation parameters found in analyzing experiments.}
\label{table:1}
\vspace{10pt}
\begin{tabular}{l|c|c|c}
\hline
Site &Distance to core, m & ${\sigma}_{R}/{\sigma}_{0}$ & cite \\
\hline
ILL & 8.76 & 0.832$\pm$ 0.075 & [19] \\
Rovno & 12 & 0.970$\pm$ 0.025 & [24] \\
Bugey-3 & 15 & 0.988$\pm$ 0.03 & [17] \\
Rovno & 18.2 &  0.995$\pm$ 0.035  & [16] \\
Rovno & 18.1 & 0.987$\pm$ 0.035  & [21] \\
SRP & 18.2 & 0.987$\pm$ 0.04 & [18] \\
SRP & 23.8 & 1.055$\pm$ 0.04 & [18] \\
Rovno & 25.2 & 1.008$\pm$ 0.049 & [16] \\
Krasnoyarsk & 33 & 1.013$\pm$ 0.051 & [22] \\
Gosgen & 37.9 & 1.018$\pm$ 0.065 & [17] \\
Bugey-3 & 40& 0.994$\pm$ 0.051 & [17] \\
Gosgen & 45.9 & 1.045$\pm$ 0.054 & [17] \\
Krasnoyarsk & 57 & 0.989$\pm$ 0.051 & [22] \\
Gosgen & 64.7 & 0.975$\pm$ 0.076 & [17] \\
Krasnoyarsk & 92 & 1.031$\pm$ 0.20 & [26] \\
Bugey-3 & 95 & 0.915$\pm$ 0.141 & [17] \\
Krasnoyarsk & 231 & 1.161$\pm$ 0.10 & [26] \\
Palo Verde & 820 & 1.011$\pm$ 0.061 & [23] \\
CHOOZ & 1050 & 0.98$\pm$ 0.043 & [24] \\
\hline
\end{tabular}\\[2pt]
\end{table} 

\section*{Conclusion}

In the article the joint analysis of reactor spectral experiments which used for neutrino registering the inverse neta decay reaction is presented. This reaction is widely used for antineutrino detection because of it high value of cross section and possibility of spectrometry.

In each particular experiment they did not find oscillations but during minimization of ${\chi}^2$ very close values of mass parameter of oscillations were found. The insufficiency of statystics did not allow to conclude exactly on presence or absence of oscillations. Proposed method of joint analysis admits to increase total statystics.

The analysis exibits two regions of probable oscillation parameters with high confidence level. These areas partially cross areas of other non reactor experiments where they announsed about oscillations obsevation.

Oscillation parameters found in the analysis ${\Delta}m^2$=0.208 eV$^2$, $\sin^{2}2\theta$ = 0.08 and ${\Delta}m^2$=0.925 eV$^2$, $\sin^{2}2\theta$ = 0.08 can be accepted with 99\% CL. Accounting that the analysis was done only for spectrum shape the mass can be taken into account as reliable. For amplitude of oscillation we can admit the high limit coming from global ratio of observable to expected flux $~0.98\pm 0.03$ $\sin^{2}2\theta \le$ 0.05.

For final conclusion on existence of neutrino oscillations with found parameters it is desirable to have some other new experiments. The experiment should have several detectors placed at cites of maximal spectrum distortion and be operatable simultaneously. This admits to exclude some undesirable uncertanties for example connected with core composition in case of nuclear reactor.

For the moment there were two proposals of so kind experiments. One is supposed to do with artifical neutrino source $^{51}$Cr and Ga target at Baksan [11]. They propose to construct special setup with two concentrical volumes filled with liquid Ga. Second one is proposed in Dubna-ITEP collaboration [27] at Kalinin NPP. They propose a new kind of neutrino detector for reactor control purposes and in parallel it might be possible to measure antineutrino spectra at distances difference of 1 m and in several detector positions.

\section*{Acknowledgments}

Author gratefully thanks L. A. Mikaelyan for useful discussions and critical notes, A.S. Starostin and L.B. Bezrukov for discussions of analysis results, interest and stimulation of work.

The calculations were made with using the compueter cluster of INR RAS.

\end{document}